\documentclass{aa}

\usepackage{graphics}  

\begin{document}

   \thesaurus{ 03.20.4;  
               08.02.7;  
               08.22.2;  
               10.07.3 47 Tuc}  

\title{SX Ph\oe nicis Stars in the Core of 47 Tucanae\thanks{Based on 
observations made with the NASA/ESA Hubble Space Telescope,
obtained at the Space Telescope Science Institute, which is operated by
the Association of Universities for Research in Astronomy Inc., under
NASA contract NAS5-26555}}


\author{H. Bruntt\inst{1} \and S. Frandsen\inst{1} 
\and R. L. Gilliland\inst{2} \and J. Christensen-Dalsgaard\inst{1,3} 
\and J. O. Petersen\inst{4} \and P. Guhathakurta\inst{5}\thanks{Alfred P. 
Sloan Research Fellow} \and P. D. Edmonds\inst{6} \and G. Bono\inst{7,8}}

   \offprints{H. Bruntt (email: bruntt@ifa.au.dk)}

   \institute{Institut for Fysik og Astronomi, 
		Aarhus Universitet, DK-8000 Aarhus C, Denmark 
         \and
		Space Telescope Science Institute, 3700 San Martin Drive,
		Baltimore, Maryland 21218, USA   
	 \and
		Teoretisk Astrofysik Center, Danmarks Grundforskningsfond,
 		Aarhus Universitet, DK-8000 Aarhus C, Denmark 
	 \and
                Astronomisk Observatorium, Niels Bohr Institutet for Astronomi, 
                Fysik og Geofysik, \\ Juliane Maries Vej 30, 
		DK-2100 K\o benhavn \O , Denmark
         \and
                UCO / Lick Observatory, University of California,
                1156 High Street, Santa Cruz, California 95064, USA 
         \and
		Center for Astrophysics, 60 Garden Street, Cambridge, 
                MA 02138, USA
	\and
		Osservatorio Astronomico di Roma, 
                via Frascati 33, 00040 Monte Porzio Catone, Italy 
	\and
		European Southern Observatory, Casilla 19001, Santiago, Chile 
}

   \date{Received; accepted }

   \maketitle

   \begin{abstract}

We present new results on five of six known SX Ph\oe nicis 
stars in the core of the globular cluster 47 Tucanae.
We give interpretations of the light curves in 
the $V$ and $I$ bands from 8.3 days of observations 
with the Hubble Space Telescope near the core of 47 Tuc. 
The most evolved SX Phe star in
the cluster is a double-mode pulsator (V2) 
and we determine its mass to be $(1.54\pm0.05) \, M_\odot$ 
from its position in the Hertzsprung-Russell diagram and by
comparing observed periods with current theoretical 
pulsation models.
For V14 we do not detect any pulsation signal. 
For the double-mode pulsators V3, V15, and V16 we cannot 
give a safe identification of the modes.
We also describe the photometric techniques we have used to
extract the light curves of stars in the crowded
core. Some of the SX Ph\oe nicis are saturated and
we demonstrate that even for stars that show signs of a bleeding 
signal we can obtain a point-to-point accuracy of 1-3\%.

   \end{abstract}

%

\section{Introduction}

Blue straggler stars (BSS) are found in all globular clusters
where thorough searches for them have been made (Bailyn \cite{bailyn}).
BSS are thought to be the results of direct collisions between 
stars or perhaps the gradual coalescence of
binary stars (Guhathakurta et al.~\cite{guha98}, Bailyn \cite{bailyn}). 
The BSS are hotter and brighter than the turn-off stars in a globular
cluster and some BSS will cross the classical instability strip for 
$\delta$ Scuti stars. The variable globular cluster 
BSS are called SX Ph{\oe}nicis (SX
Phe) stars, using the name for the prototype population II field star.

\begin{table}
\caption[ ]{Fundamental parameters for 47 Tucanae. Values are taken from
VandenBerg (\cite{vandenberg2000})}

\label{frames}

\vskip 0.2cm

\centering{
\begin{tabular}{rrrrr}  \hline
\noalign{\smallskip}

[Fe/H]  & $E(B-V)$  & Age [Gyr]    & $ V_{TO}$ & $(m-M)_V$ \\ 
\hline
\noalign{\smallskip}
$  -0.83$ & $   0.032$ & $    11.5 $ & $  17.65$ & $  13.37$ \\ 
$\pm0.10$ & $\pm0.005$ & $  \pm0.8 $ & $\pm0.10$ & $\pm0.05$ \\ 

\hline
\noalign{\smallskip}

\end{tabular}}
\end{table}

When discussing globular clusters it is important to realize
that they are not just a large group of individual independent stars. 
BSS are but one piece of evidence for non-standard stellar evolution due to
dynamical processes (Camilo et al.~\cite{pulsars47}, Bailyn \cite{bailyn}). 
For the past five years a number of globular cluster cores have
been probed to look more closely at the populations of stars that are
direct evidence that the dynamics of stars have influence on the 
evolution of globular clusters as a whole.
This is possible with the Hubble Space Telescope (HST) and
has indeed yielded some interesting surprises: 
1) BSS are found in all globular clusters (Bailyn \cite{bailyn}). 
2) The horizontal-branch morphology possibly
depends on the dynamical evolution of the cluster, although
the metallicity is the most important parameter 
(Fusi Pecci et al.~\cite{horizbranch}).
3) The discovery of cataclysmic variables in NGC 6397 
(Cool et al.~\cite{cool}).

BSS and in turn SX Phe stars are found primarily 
near the cluster core as a result 
of mass segregation (see e.g.~Edmonds et al.~\cite{edmonds}).
With the exceptional resolving power of HST 
it is possible to probe the cores 
of globular clusters, examples of such research are
47 Tuc (Gilliland et al.~\cite{gill98}) and M5 
(Drissen \& Shara \cite{drissen}) --- but see also 
Piotto et al.~(\cite{glob-cores}).

\subsection{\label{sec:var}Variable stars in 47 Tuc}

The globular cluster 47 Tucanae is one of several clusters which have been 
searched for binaries and variable stars with the HST. Early results
showed that 47 Tuc contained a significant number of 
BSS (Guhathakurta et al.~\cite{guha92}).
With observations carried out with the uncorrected optics of HST 
(September 1993) a study of variable 
stars in 47 Tuc was made by Edmonds et al.~(\cite{edmonds}).
In this search a few SX Phe and several
eclipsing binaries were found (Edmonds et al.~\cite{edmonds})
as well as variable K giant stars (Edmonds \& Gilliland \cite{kgiants}).

In particular the BSS were analyzed by Gilliland et al.~(\cite{gill98}) 
who found six SX Phe stars.
{}From the location of the double-mode variables in the {\em Petersen diagram} 
(Petersen \& Christensen-Dalsgaard \cite{otzen96})
Gilliland et al.~(\cite{gill98}) were able to estimate the 
{\em pulsation masses} for the four double-mode SX Phe stars.
Indeed, they showed that the pulsation masses and the 
{\em evolutionary masses}, which were estimated from 
the positions in the HR diagram,  
agreed quite well. Combining these two methods of {\em weighing} stars, 
they found masses from $(1.35\pm0.1)$ to $(1.6\pm0.2) \, M_\odot$ well above
the turn-off mass at $\approx0.85 \, M_\odot$ in agreement with the generally
accepted merging or colliding star scenario for BSS
(Bailyn \cite{bailyn}).

\subsection{Asteroseismology of SX Phe stars}

Among the SX Phe stars the double-mode pulsators are particularly interesting.
In some cases it is possible to identify the pulsation modes.
This class of star is comparable to the classical double-mode Cepheids 
(Petersen \& Christensen-Dalsgaard \cite{otzen96}).
The SX Phe stars are often 
pulsating in the fundamental mode and the first overtone or perhaps 
in two modes of higher overtones (second and third, third and fourth, etc.). 
Detections of double-mode SX Phe oscillation have been made in 47 Tuc
(Gilliland et al.~\cite{gill95,gill98}), $\omega$ Cen (Freyhammer et al.~\cite{omegacen}), 
and in a few field variables including SX Phe itself
(e.g.~Garrido \& Rodriguez \cite{garrido}).

The observed period ratio and the period of the main mode depend quite
sensitively on the mass, metallicity, and evolutionary stage of the star
(i.e.~the age for a given mass). 
The ratio of the periods of the double-mode pulsators
are well determined from theoretical models:
As discussed in Section \ref{sec:dmvtwo} below, the intrinsic precision of
the theoretical period ratio is better than
$0.01$ percent (see also Petersen \& Christensen-Dalsgaard \cite{otzen96}). 
Hence it is certainly possible
to infer physical properties of the double-mode SX Phe stars.

There is evidence that some observed SX Phe stars oscillate in 
non-radial modes 
and it may become possible to extract information about the properties
of the cores of these stars.
Examples of observed SX Phe stars that show evidence of
non-radial modes are SX Phe itself (Garrido \& Rodriguez \cite{garrido}) and 
V3 in 47 Tuc (Gilliland et al.~\cite{gill98}).
Progress in the understanding of SX Phe stars 
can also be made if one can detect low-amplitude modes,
which may provide more detailed information about the internal structure 
of these stars. 

%
%

Section \ref{data} contains a description of the data set.
In Section \ref{reduction} we discuss the reduction of the 
data including the task of dealing with saturated bleeding pixels on the CCD.
Section \ref{analysis} contains the details of the time-series analysis while
in Section \ref{results} we present the results with emphasis on 
the double-mode SX Phe stars. Finally, Section 6 summarizes our conclusions.

\begin{figure}
\resizebox{8.8cm}{!}
{\includegraphics{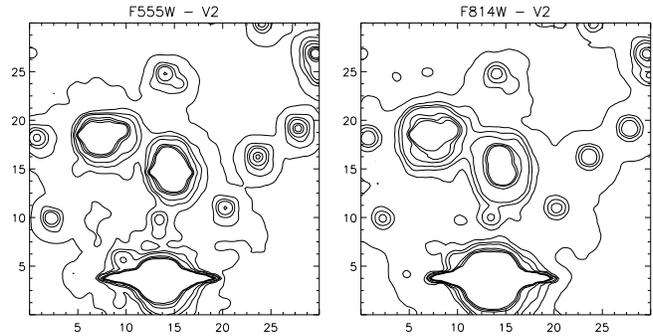}}
\caption{Contour plots around the SX Phe star V2 in the super frame with 
the F555W and F814W filter. V2 is at position (15,15) and is clearly
contaminated by another bright star above it (15,16). In F555W it can be seen
that both stars are {\em bleeding} in the horizontal (CCD read-out) direction}
\label{fig:bleeder}
\end{figure}

\section{\label{data}Observations}
In the present work we have analyzed an 8.3 day time series of the
stars in the central part of 47 Tuc.
The data were obtained with WFPC2 on the HST in July 1999. 
The main goal of the project
was to search for giant planets around the main-sequence stars 
of the cluster (Gilliland et al.~\cite{gill2000}).
The primary exposure times were 160 s in F555W and F814W filters 
($V$ and $I$\footnote{$I$ magnitudes refer to the Cousins filter}) and
optimized for stars at and below the turn-off. Consequently, the luminous
BSS stars are all saturated in $V$ and some are also saturated in $I$.

The PC1 CCD of WFPC2 was placed near the center of 47 Tuc. This is 
where the density of stars is highest and most of the BSS stars are found.
Five of the six known SX Phe stars in the core of 47 Tuc 
(Gilliland et al.~\cite{gill98}) are within the field of view 
(V1 fell outside).

The present study is an improvement compared to the results
by Gilliland et al.~(\cite{gill98}). They used HST photometry (with the
aberrated WF/PC) in the $U$ band with a total coverage of 39 hours. 
Although ideal for the detection of stellar
oscillations the exposure time was 1000 s compared to 160 s
for the present data set. 
The new time series are more than five times longer with total of 1289
data points compared to the 99 points in the old data set. 
Although the amplitudes are about a factor
of two smaller (using $V$ and $I$ instead of $U$) 
for the present data set and
some of the stars are saturated, we expect to be able to 
derive more accurate periods for the SX Phe stars.

\section{\label{reduction}Data reduction}

We have used the DAOPHOT/ALLSTAR (D-A) software  
which is designed for doing photometry in crowded fields (Stetson \cite{daophot-art}). 
For this study we have only reduced the PC1 frames as the BSS stars are 
found near the core of 47 Tuc. We use D-A to construct an empirical PSF model 
from about 60 stars. 
Because of the extreme crowding near the core of 47 Tuc this is a difficult task.
The PSF model is iteratively improved by 
cleaning the neighbouring stars around the stars used for 
creating the PSF model.

It is well known that the PSF changes with time due to the 
{\em breathing} of HST (Suchkov \& Casertano \cite{breathing}). 
We also need a good description of the PSF 
wings as this has serious impact on the accuracy of the 
photometry of the saturated SX Phe stars: For these stars the light in the
core is not used in the D-A analysis, and we have to look for oscillations in the wings 
of the PSF profile.
Thus we constructed empirical PSF models for all frames. 

\begin{figure}
\resizebox{8.8cm}{!}
{\includegraphics{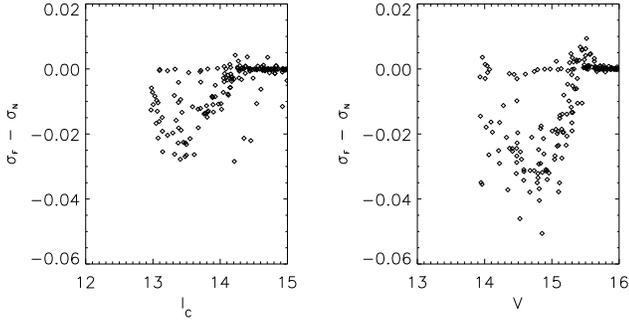}}
\caption{Difference in internal standard deviation of light curves for
two different reduction techniques for the $I$~(left) and $V$~(right)
filters. A significant improvement is obtained for the brightest stars}
\label{fig:improve}
\end{figure}

To improve the photometry we have stacked all the 
frames in each filter to create a {\em super frame}. 
This technique is described in Gilliland et al.~(\cite{gill99}).
In this way we are able to combine the information from all frames to obtain
accurate positions of the stars in each individual frame. 
This is done by calculating
the offset position from the stars in the super frame to each individual frame 
from the position of several hundred reasonably isolated and bright stars. 
In order to be able to account for geometric distortion, 
we find the six coefficients ($a_1$ - $a_6$) in the transformation
$$X_i=a_1 + a_2\, x_i + a_3\, y_i + a_4 \, x_i^2 + a_5\, y_i^2 + a_6\, x_i \, y_i$$  
for each frame; here ($x,y$) is the position in the super frame for star $i$,
and $X$ is the horizontal position on a given frame. 
The same is done for the $Y$ coordinate.

We then redo the D-A photometry with the improved positions 
of the stars being {\em fixed}; thus the
magnitudes are the only parameters which are fitted by D-A. 
This step greatly improves the photometric accuracy.

\begin{figure}
\resizebox{8.8cm}{!}
{\includegraphics{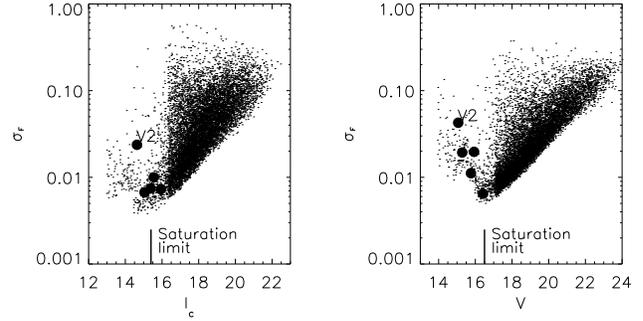}}

\caption{The final internal standard deviation of 
light curves for the $I$ (left) and $V$ (right) filters. 
The SX Phe stars have been clearly marked}
\label{fig:stdev}
\end{figure}

\subsection{\label{sec:bleeder}Bleeding pixels}

We find that the photometry of the brightest saturated stars is quite poor. 
By inspecting contour levels around stars with different degrees of saturation 
it is obvious that the signal from the saturated core starts 
to {\em bleed} along the read-out direction. Thus the stellar profile
is slightly elongated, i.e.~the nice symmetry of the PSF
is broken due to the contamination from the bleeding pixels. 
A specific example of bleeding stars is shown in Figure \ref{fig:bleeder}.
Note that the data analyzed here have been rotated by 90 degrees from 
normal WFPC2 conventions, hence bleeding along $x$.
We assume that the signal in pixels that are neighbours to saturated cores
may not be reliable. To correct for this we used different methods to find
the pixels that were affected. 
It turned out that a simple correction gave good results:
We have found that the contamination due to bleeding pixels seems 
to set in at a certain degree of saturation, 
i.e.~stars for which only the central one or two pixels are 
saturated {\em do not} suffer 
from contamination. The best result was obtained by {\em flagging} pixels 
which are neighbours to three or more pixels (in the read-out direction) 
that are above the saturation limit. Flagging simply means that the pixel value
is set to a value above the saturation limit, and is hence treated as 
such by D-A, i.e.~it is not used for the PSF fitting and hence 
the determination of the magnitude.

The improvement for the saturated stars is significant. 
In Figure \ref{fig:improve}
we show the difference in internal standard deviation (ISD) for 
the light curves
of all stars on PC1 before (N) and after flagging (F) the contaminated pixels. 
The final ISD for all stars is shown in Figure \ref{fig:stdev} in which
the five SX Phe stars have been emphasized. We define
$$\sigma_{\rm ISD}^2 = {\sum_{i}^{N-1} (m_ i - m_{i+1})^2} / [2(N-1)] \; ,$$
where $N$ is the number of observations.

\begin{figure}
\resizebox{8.5cm}{!}
{\includegraphics{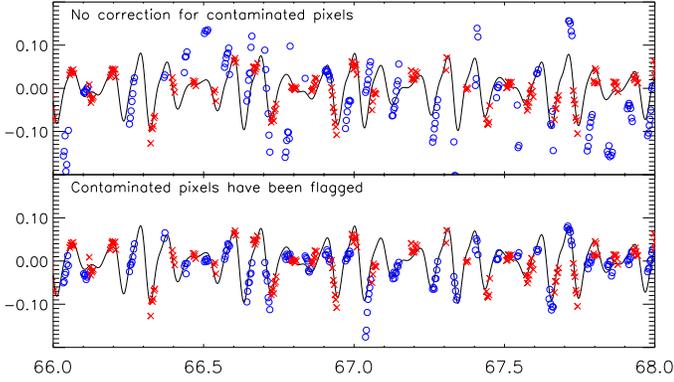}}
\caption{Light curve for SX Phe star V2 in F555W ($\circ$) and 
F814W ($\times$) when using two different reduction techniques 
(see text for details). The abscissa is the heliocentric Julian 
date minus 2451300}
\label{fig:improve-v2}
\end{figure}

A specific example of the importance of flagging the contaminated 
pixels is shown
in Figure \ref{fig:improve-v2}. This shows the light curve in both 
F555W ($\circ$) 
and F814W ($\times$) before and after the flagging. 
Due to the
amplitude dependence on wavelength the F555W light curve has been scaled by an 
empirical factor of $A_{814}/A_{555}=0.639$ 
(this scaling is also made before the time-series analysis in
Section \ref{analysis}).

\subsection{\label{sec:gill-analysis}Further analysis}

Two other time-series analyses should be mentioned here.  
Gilliland et al.~(\cite{gill2000}) describe 
use of difference image analyses for obtaining optimal
results for the unsaturated stars comprising the primary data use.
The difference image analysis provided time-series precisions averaging
a factor of two better than those presented here for unsaturated stars,
but cannot be applied at all to the saturated stars. A second approach
to saturated-star extractions has been developed by one of us (RLG,
Gilliland \cite{gill94}) making use of aperture photometry including 
the saturated pixels. The latter approach provides generally comparable 
results to those described here for the PC1 extractions.

\section{\label{analysis}Time-series analysis}

\begin{figure}
\resizebox{8.8cm}{!}
{\includegraphics{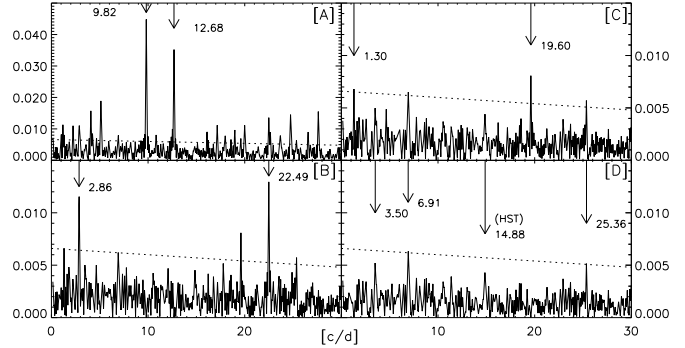}}
\caption{Amplitude spectra 
for the SX Phe star V2 computed 
when using the combined F555W and F814W light curve 
(top left plot and labeled A).
The other three plots (labeled B,C, and D)
show the resulting amplitude spectra when
the modes marked by arrows have been subtracted. 
Note that the scales on the ordinate differ. 
The dotted line is the 4$\sigma$ detection limit}
\label{fig:power-v2}
\end{figure}

To perform the frequency analysis of the light curves
we used the software package {\em period98} developed by
Sperl (\cite{sperl98}). Figure \ref{fig:power-v2} shows the
amplitude spectrum for the SX Phe star V2 (top left plot) and 
also the resulting amplitude spectra when
the variation corresponding to the most significant peaks has been subtracted.
The arrows show the frequencies in cycles per day (c/d) 
--- one barely signifi\-cant 
peak at 14.88 c/d is labeled ``HST'', and may be due to the
96.4 minutes orbital period of the observatory. The dotted 
line is the empirically 
determined 4$\sigma$ detection limit; peaks below this are 
not considered to be significant. All mode frequencies,
amplitudes, phases, and errors are quoted in Table \ref{tab:periods}.
We note that the light curves have been analyzed
independently by three of the authors using different software --- and
our results agree.

\begin{figure}
\resizebox{8.8cm}{!}
{\includegraphics{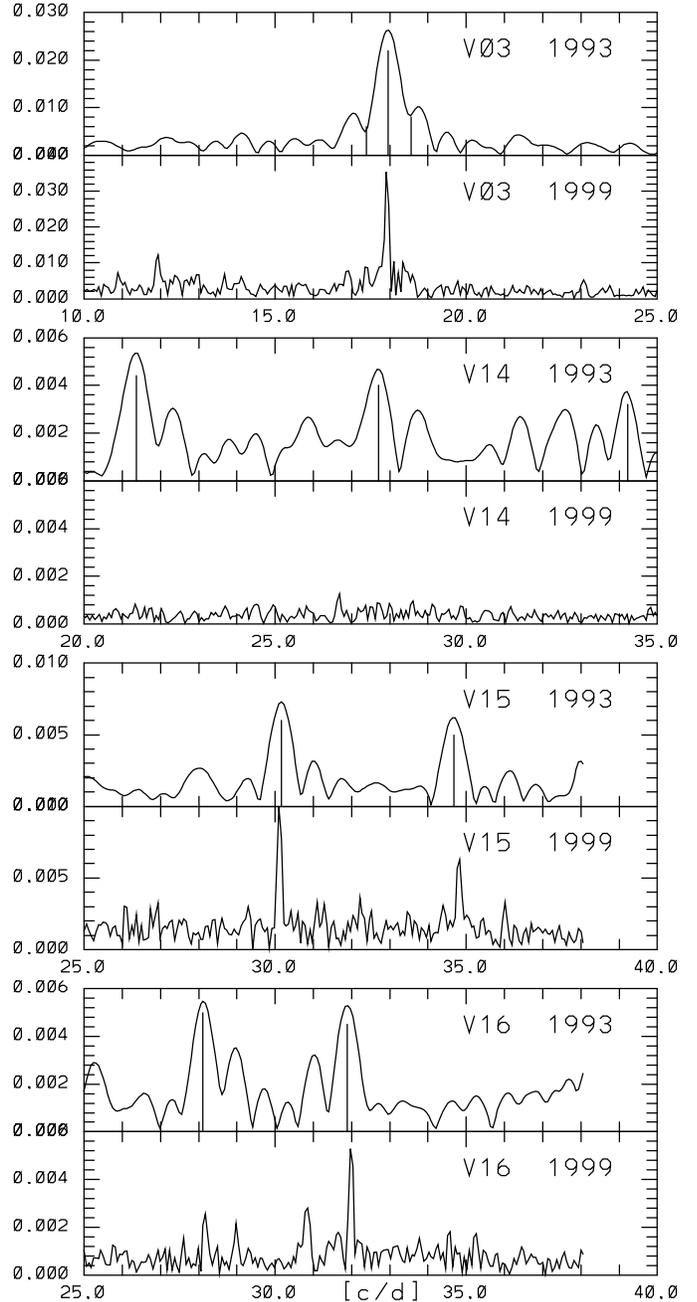}}
\caption{Amplitude spectra for V3, V14, V15, and V16 are shown comparing 
the 1993 (Gilliland et al.~ \cite{gill98}) results with the new 1999 data.
All amplitudes have been scaled as needed by wavelength to provide
V-band equivalent magnitudes.  Frequency units are cycles per day.
All modes claimed from the 1993 data are marked 
with vertical bars under the corresponding peaks}
\label{fig:powerALL}
\end{figure}

The primary modes of V2 have sufficient amplitude to provide unambiguous
detections from both the 1993 (Gilliland et al.~ \cite{gill98}) and 1999
time series.
We now consider the other four SX Phe stars for which some of the modes
claimed previously were near the detection limit provided by the 1993 data.
Figure \ref{fig:powerALL} shows the amplitude spectra for V3, V14, V15,
and V16 with the original $U$-band (1993) and $I$-band (1999) amplitudes
transformed to a common $V$ estimate by scaling by central wavelength ratios.
Time series for V3 and V15 are based upon combined $V$ and $I$ (as for V2)
data making use of techniques previously described for analyzing saturated stars.
For V14 and V16 the stars never saturate in $I$ exposures; for these
we use only the 653 unsaturated images and the more precise photometry
provided by difference image analyses.
Results for V3, V15, and V16 show that all primary modes claimed from the
1993 data are present, and these will be further discussed in Section 5.5.
V14 was the least significant detection from the 1993 data, and was only
selected (Gilliland et al.~ \cite{gill98}) as a likely variable by searching
for repeated low amplitude peaks with spacing characteristic of successive
radial modes.
The 1999 data for V14 clearly show that not even a slight hint of the 
claimed modes exists; the most likely interpretation is that V14 is not 
a variable and we will not further consider V14 as an SX Phe star.
The V14 amplitude spectrum for 1999 nicely illustrates a characteristic of the 
extensive $HST$ observations -- a very clean spectrum that in fact stays
flat from below 1 cycle/day out to the Nyquist frequency of 180 cycles/day.
V16 was also only found as a variable from the 1993 data based on a search
for successive radial overtone peaks; in this case the 1999 data clearly
confirm oscillations are again present.

\begin{table*}[hbtp]
\caption{\label{tab:periods}Frequencies, amplitudes, and
phases for modes for the observed SX Phe stars. 
Modes above the 4$\sigma$ level are marked with $*$ while
less significant modes are marked with $?$.
Numbers in parenthesis are the statistical errors. 
The frequencies found 
by Gilliland et al.~(\cite{gill98}) are given in the rightmost column.
Note that the formal errors on the frequencies of the new results
are lower by almost an order of magnitude compared to 
Gilliland et al.~(\cite{gill98}). 
In column 5 $\nu_1$ and $\nu_2$ are the frequencies of the primary and the
secondary mode}
\centerline{\begin{tabular}{crlclr}\hline
Star  &Frequency  &$V$ Amplitude   &Phase        & Mode   & Gilliland et al.\\
      & [c/d]     & [mag]      & [c]         & combination & \\ \hline  
V2 
 &  {\bf 9.816}(0.002)&{\bf0.044}(0.001)&{\bf0.662}(0.004) &* $\nu_1$ &{\bf  9.807}(0.013)\\
 &  {\bf12.684}(0.002)&{\bf0.034}(0.001)&{\bf0.488}(0.005) &* $\nu_2$ &{\bf 12.700}(0.016) \\
 &  {\bf22.490}(0.005)&{\bf0.012}(0.001)&{\bf0.187}(0.014) &* $\nu_1+\nu_2$ &{\bf 22.494}(0.057) \\
 &  {\bf 2.866}(0.007)&{\bf0.011}(0.001)&{\bf0.620}(0.014) &* $\nu_2-\nu_1$ & \\
 &  {\bf19.604}(0.020)&{\bf0.008}(0.001)&{\bf0.564}(0.076) &* $2\nu_1$ & \\
 &  {\bf25.354}(0.013)&{\bf0.005}(0.001)&{\bf0.351}(0.035) &? $2\nu_2$ & \\
 &  {\bf 6.914}(0.013)&{\bf0.005}(0.001)&{\bf0.179}(0.025) &? $2\nu_1-\nu_2$ & \\
\hline V3 
 &  {\bf17.929}(0.001)&{\bf0.036}(0.005)&{\bf0.624}(0.007)   &* $\nu_1$ &{\bf 17.948}(0.022) \\
 &  {\bf18.359}(0.004)&{\bf0.009}(0.005)&{\bf0.925}(0.017)   &? $\nu_2$ &{\bf 18.585}(0.064) \\
\hline V15 
 &  {\bf30.110}(0.004)&{\bf0.0084}(0.0004)&{\bf0.727}(0.007)&* $\nu_1$  &{\bf 30.157}(0.053)  \\
 &  {\bf34.807}(0.010)&{\bf0.0038}(0.0004)&{\bf0.493}(0.016)&* $\nu_2$  &{\bf 34.671}(0.070)\\
\hline V16 
 &  {\bf31.988}(0.005)&{\bf0.0039}(0.0005)&{\bf0.055}(0.016)&* $\nu_1$  &{\bf 31.894}(0.064) \\
 &  {\bf28.165}(0.008)&{\bf0.0016}(0.0005)&{\bf0.256}(0.016)&? $\nu_2$  &{\bf 28.116}(0.071) \\
\hline
\end{tabular}}
\hspace{1mm}

\end{table*}

\begin{figure}
\resizebox{8.8cm}{!}
{\includegraphics{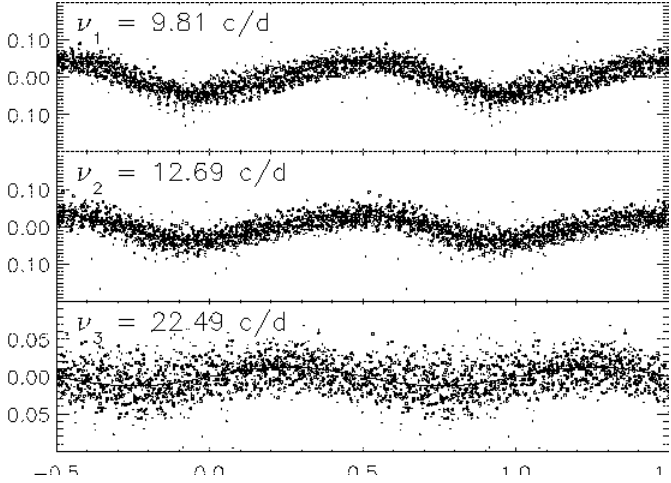}}
\vskip 0.1cm
\caption{Phase diagrams for V2 for the three main modes}
\label{fig:phV2}
\end{figure}

\begin{figure*}
\hskip 0.2cm
\resizebox{17.2cm}{!}
{\includegraphics{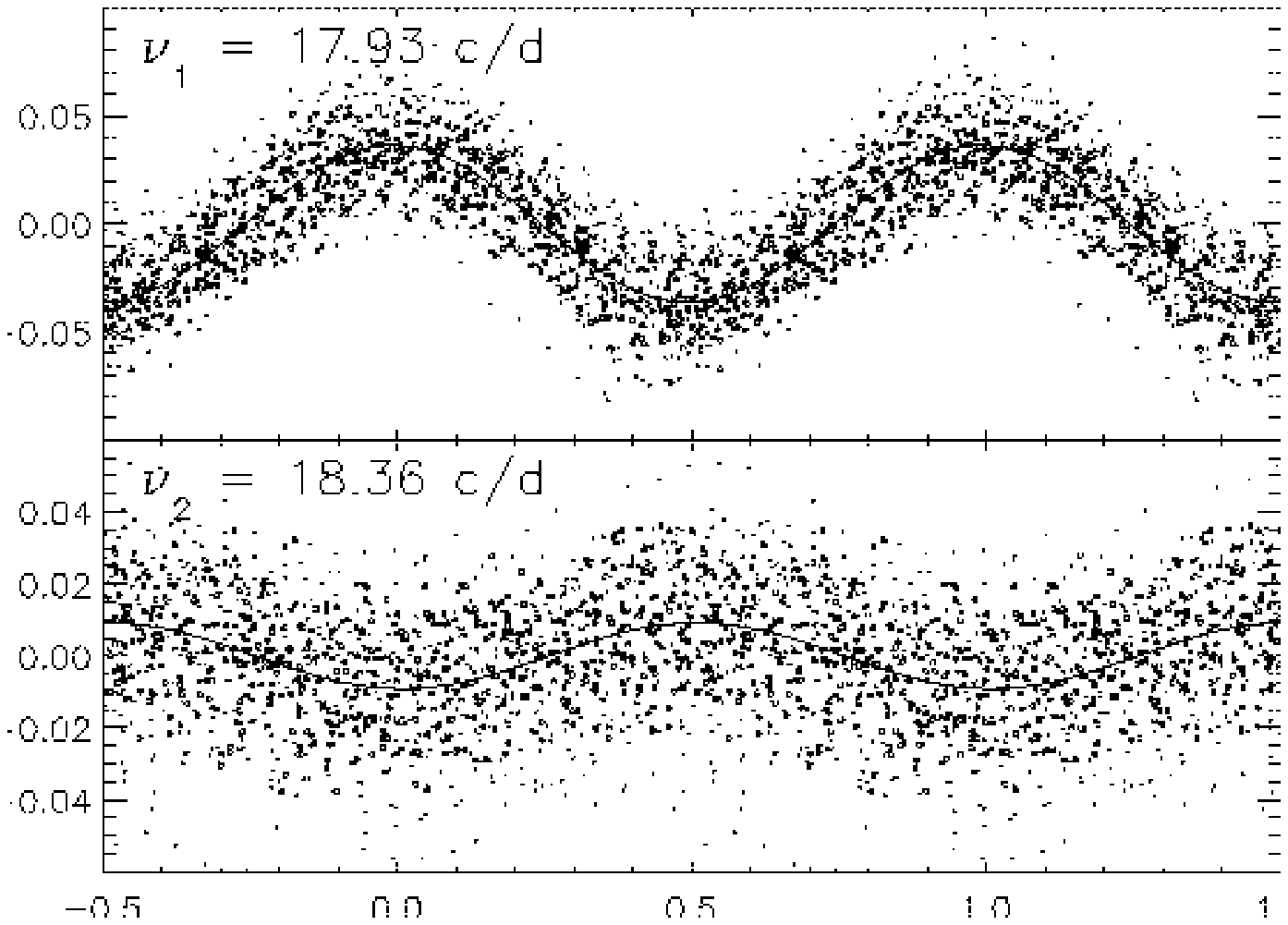} 
\hskip 0.4cm
\includegraphics{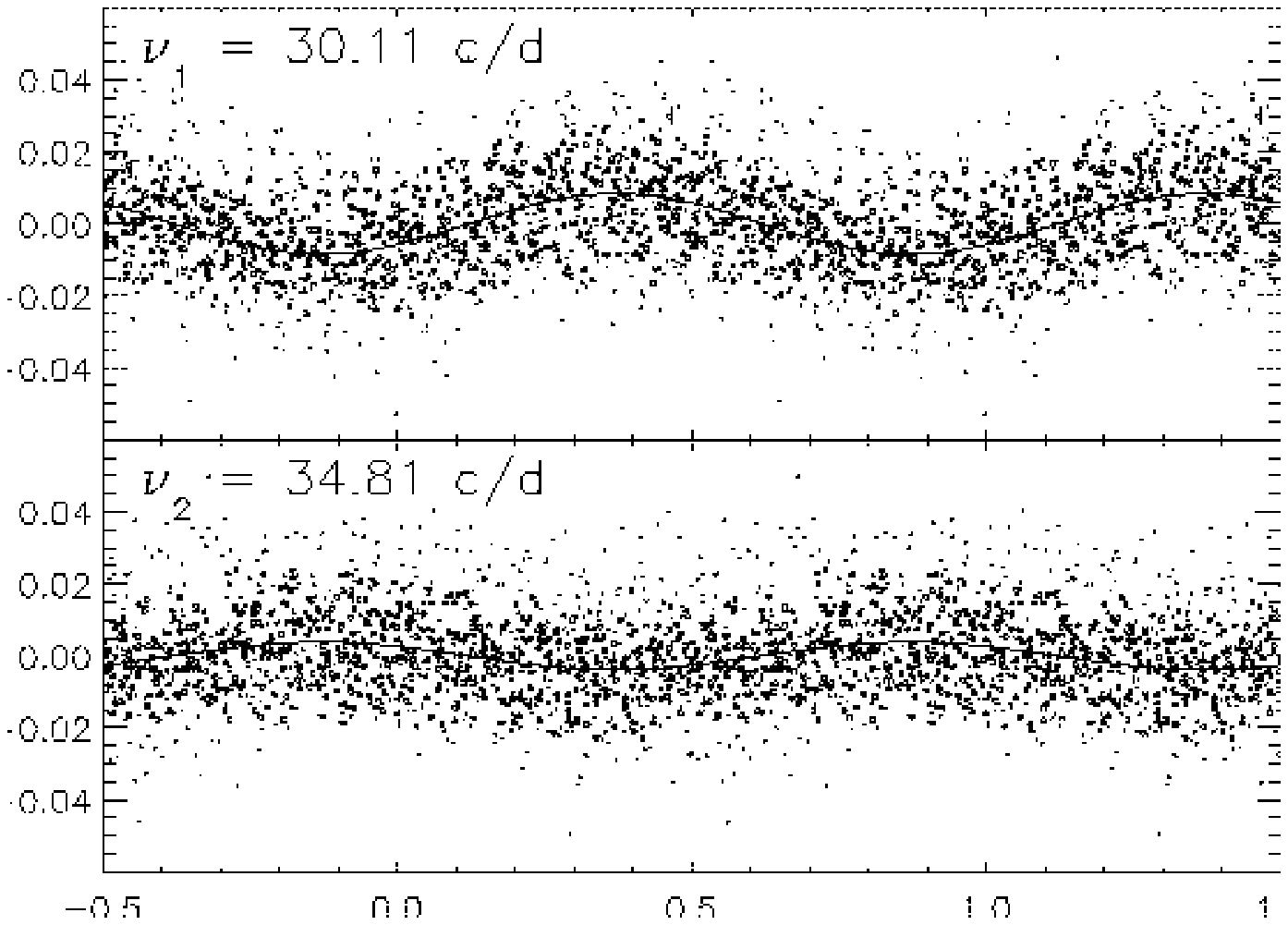}}
\vskip 0.1cm
\caption{Phase diagram for V3 and V15 for two modes}
\label{fig:phALL}
\end{figure*}

In Figures \ref{fig:phV2} and \ref{fig:phALL} we show the 
phase diagrams for the three largest amplitude 
double-mode stars in our sample. In each diagram all known modes have
been subtracted except for one mode, the frequency of which is shown
in the top left corner. Note the different scale on the ordinates.

\subsection{Search for new SX Phe stars}

  The original HST data 
  (WF/PC; Gilliland et al.~\cite{gill98}) used to find 
  SX Phe variables in 47 Tuc covered an area of 66 $\times$ 66 arcsec$^2$, 
  with the cluster centre about 10 arcsec from the field of view centre.
  The new search domain with WFPC2 observing in 1999 is such that 
  the PC1 field of view is almost entirely contained within the 
  original search area, while the much larger field of view of the 
  WF CCDs is (at the 90\% level) outside the original search space.
  We have extracted time series for 2968 saturated stars on the 4 
  WFPC2 CCDs using the technique of Gilliland (\cite{gill94}). 
  Near the domain in the CMD where the previous SX Phe stars were found 
  the root-mean-square variations in the resulting time series are
  $\sim$ 1 -- 2\% (albeit with non-Gaussian
  characteristics).  We note that the previously detected SX Phe 
  variables V2, V3, V15, and V16 are easily observed in these new 
  time series.
  Although the new survey field of view is much larger than that 
  previously surveyed, it is also on average farther from the cluster
  center.  In the domain $15.0 < V < 17.0$, $0.2 < V - I < 0.45$
  the original search contained about 25 stars of which 6
  are SX Phe variables.  The number of new stars surveyed within this
  box in the CMD is only 7 consistent with the known, strong central
  concentration of BSS.  The time series for all of the 
  newly surveyed BSS were analyzed by taking power spectra
  and searching for significant peaks; no convincing evidence of 
  variables was found.  Amplitudes of 0.01 magnitudes would in 
  general be easily seen unless the period is similar to the 96.4 minute
  HST orbital period or its harmonics.

\section{\label{results}Results}

\subsection{\label{sec:cmd}The colour-magnitude diagram}

The conversion from 
the instrumental to standard $V,I$ (Johnson-Cousins) magnitudes was
done following the ``photometric cookbook'' of Holtzman 
et al.~(\cite{holtzman}). We correct for geometric distortion 
and the ``34th row effect'' (Anderson \& King \cite{34throw}).
We have corrected for the charge-transfer 
efficiency (CTE) problem by following the formulation 
of Stetson (\cite{cte-effect}) and we include his new 
photometric zero-points.

Systematic differences between long and short exposures
obtained with the HST have been reported in the literature 
(see e.g.~Kelson et al.~\cite{longshort}). There is a
tendency that measurements of stellar brightness 
from frames with long exposure times
yield brighter magnitudes than short-time exposures.
To be able to correct for the ``long-short anomaly'' we reduced some
additional short exposures obtained for our field: Four 1.6 s exposures in
F555W and F814W (the same field as the 160 s exposures).

We then determined 
the offsets in $V$ and $I$ from frames with different exposure 
times. The offsets are: 
$\Delta V$(160-1.6s)$ = -0.104\pm0.004$ (1042 stars), and 
$\Delta I$(160-1.6s)$ = -0.0554\pm0.0035$ (1034 stars).
We stress that this is after correcting for the CTE effect, i.e.~the
long-short anomaly is ``real'' and not only a part of the CTE problem.
A possible explanation for the offsets is the difficulty of 
determining the $0.5$ arcsec
aperture correction due to crowding.
This correction is needed when following the calibration ``cookbook''
of Holtzman et al.~(\cite{holtzman}).

After the removal of the offsets 
we compare our CMD with several results from the literature.
We have found the average $V$ of the stars in the transition between
the turn-off and the red giant branch:
$V_{\rm TO/RGB} = 17.31$ near $V-I=0.80$. 
The mean of Hesser et al.~(\cite{hesser}), Kaluzny et al.~(\cite{kaluzny}), 
and Alcaino \& Liller (\cite{alcaino}) is $V_{\rm TO/RGB} = 17.18 \pm 0.04$. 
We thus apply an offset of $\Delta V = 17.18 - 17.31 = -0.13$. 
The offset in $I$
is found from the colour of the turn-off. The mean colour from Kaluzny et al. (\cite{kaluzny})
and Alcaino \& Liller (\cite{alcaino}) is $(V-I)_{\rm TO} = 0.695 \pm 0.02$.
We find  $\Delta I = -0.096$. 

\begin{figure}
\hspace{-.7cm}
\resizebox{9.8cm}{!}
{\includegraphics{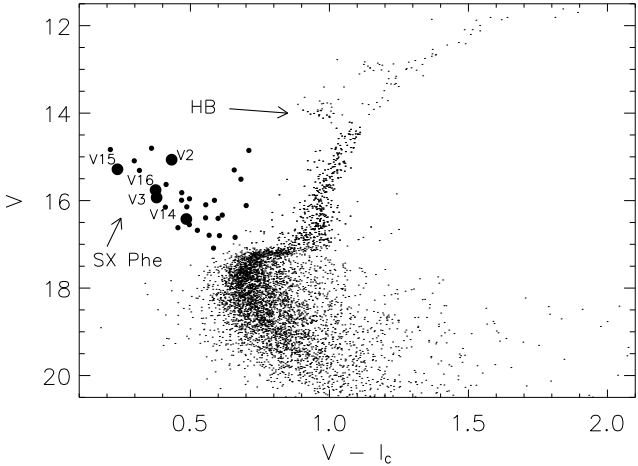}}
\caption{Combined CMD for 47 Tuc.
Filled dots mark the BSS stars, and the SX Phe stars have
been labeled}

\label{fig:cmd}
\end{figure}

The final colour-magnitude diagram (CMD) is shown in Figure \ref{fig:cmd} 
where the BSS stars have been marked. The parameters of the SX Phe
stars are given in Table \ref{tab:params}.

Two groups in our team have performed the
standard photometry in order to detect systematic differences.
The first group reduced each short exposure image individually while the second
group worked on combined images.
The photometry of the SX Phe stars and the CMD presented here are
the results of the first group --- here we mention the main differences.
After correcting for the offsets mentioned above 
(TO/RGB transition level and colour of the turn-off) the results
from both groups based on the short exposures generally agree:
We note that for the CMD presented here, the brightest 
part ($V<14.5$) of the giant branch is more red ($\Delta (V-I)=0.10$) 
than the ground-based study by Da Costa \& Armandroff (\cite{redbranch}).
Our second group finds $\Delta (V-I)=-0.05$ for these stars.
In the interesting case of the SX Phe star
V2 both groups agree that $V=15.07\pm0.02$ while 
we find discrepant results for the red filter: $I=14.63\pm0.03$
and $I=14.80\pm0.03$. The most immediate explanations for this 
are the crowding around V2 (cf.~Figure \ref{fig:bleeder})
and that the images are under-sampled.
We note that systematic errors of the order 0.03 magnitudes 
may still be present (see e.g.~Stetson \cite{cte-effect}) for all stars.

To transform the empirical data into the theoretical plane we adopted 
the bolometric corrections (BCs) and the colour-temperature relations 
provided by Bessell, Castelli, \& Plez (1998)\footnote{
http://kurucz.harvard.edu/grids.html, no overshooting models}. 
By means of a linear interpolation of the tables for ${\rm [Fe/H]}=-0.5$ and 
$-1.0$ we estimated a new list of BCs for ${\rm [Fe/H]}=-0.83$.   
For MS stars we use $\log g=4.5$, while for BSS we use $\log g=4.0$. 
As a special case we assumed $\log g=3.7$ for V2, since this SX Phe has 
already evolved off the MS. Note that such a value is somewhat 
lower than the value obtained on the basis of an evolutionary track 
with $M/M_\odot=1.5$ (VandenBerg, private communication). 
On the basis of these assumptions and of the observed colour (two
independent determinations) we find that 
the effective temperature for V2 ranges 
from $T_{\rm eff}=7100$~K ($(V-I)_0=0.39$) 
to $T_{\rm eff}=7700$~K($(V-I)_0=0.23$). 
We have used $T_{\rm eff}=7100$~K for V2 and find BC $=-0.02$ mag. 
The BCs for all stars were estimated by adopting a 
similar procedure. The BC for the SX Phe stars are listed in Table 3.

The distance modulus of 47 Tuc has been determined several times
in the literature.
VandenBerg (\cite{vandenberg2000}) found $(m-M)_V = 13.37\pm0.05$ (using
new isochrones and $B,V$ photometry by Hesser et al.~\cite{hesser}).
Grundahl et al.~(\cite{grundahl}) made Str\"omgren 
photometry on 47 Tuc and 14 field sub dwarfs.
By fitting the lower main sequence of 47 Tuc
to the field stars they find a distance modulus 
$(m-M)_V = 13.32\pm0.04$. The quoted errors ignore systematic
errors, e.g.~the zero point of the metallicity scale 
and the error on the interstellar reddening. 
The systematic error is of the order 0.10 magnitudes.
These three independent studies all seem to deviate significantly from 
Carretta et al.~(\cite{hipparcos-dm}) 
and Salaris \& Weiss (\cite{salaris}) who find 
$(m-M)_V = 13.55\pm0.09$ and $13.50\pm0.05$, respectively. 
We finally draw attention to
the study of the white dwarf cooling sequence by 
Zoccali et al.~(\cite{zoccali99}). They find a distance modulus 
of $(m-M)_V=13.15\pm0.12$ (assuming again $E(B-V)=0.032$). 
In the following, we adopt the value obtained by
VandenBerg (\cite{vandenberg2000}), i.e.~$(m-M)_V=13.37\pm0.05$.

\subsection{\label{sec:hr}The Hertzsprung-Russell diagram}

Figure \ref{fig:jcdmodel} 
is the Hertzsprung-Russell (HR) diagram for 47 Tuc 
in which all BSS stars and main-sequence stars with the best 
photometry are shown. The five known SX~Phe stars have been 
labelled.
Also seen in the plot are evolutionary tracks
for models computed with the evolution code by 
Christensen-Dalsgaard 
(Petersen \& Christensen-Dalsgaard \cite{otzen96,otzen99}). 
The abundances of hydrogen, helium, and heavy elements
are $(X,Y,Z)=(0.754,0.241,0.005)$ which is appropriate for 47 Tuc
according to VandenBerg (\cite{vandenberg2000}).

We note that Salaris \& Weiss (\cite{salaris}) discuss the evidence
for a somewhat higher value of $Y=0.273$ for 47~Tuc. For example
they use the $R$-method (Buzzoni et al.~\cite{buzzoni}) with
$R=1.75\pm0.21$ and $1.86\pm0.36$ based on
Buzzoni et al.~(\cite{buzzoni}) and Hesser et al.~(\cite{hesser}),
respectively. This corresponds to a mean $Y=0.28\pm0.04$ which is 
roughly consistent with the value we have adopted (using the 
calibration by Buzzoni et al.~\cite{buzzoni}).
We note that a more recent determination of the $R$ parameter was
done by Zoccali et al.~(\cite{zoccali00}) based on observations 
with HST. They find $R = 1.52\pm0.13$ corresponding to $Y=0.246\pm0.012$ 
(again using Buzzoni et al.~\cite{buzzoni}). 
Based on this short discussion we have adopted the same 
helium content as VandenBerg (\cite{vandenberg2000}), i.e.~$Y=0.241$.

The age of the cluster was determined to be 
11.5 Gyr by VandenBerg (\cite{vandenberg2000}) from the $B,V$ photometry
by Hesser et al.~(\cite{hesser}). The new combined $V,I$ photometry from HST 
indicates that this age estimate may be a bit too young. 
In Figure~\ref{fig:jcdmodel} we have also shown an isochrone of age
12.5 Gyr computed by VandenBerg (private communication).
For comparison the $0.9 \, M_\odot$ evolutionary model 
in Figure~\ref{fig:jcdmodel} reaches the 
turn-off at $\log T_{\rm eff}=3.78$ after $\approx 10$ Gyr and
starts climbing the Hayashi-track around $\approx 13$ Gyr: This evolutionary
model is indeed comparable with stars near the observed turn-off.
Thus our models are in qualitative agreement with VandenBerg.
We note that Salaris \& Weiss (\cite{salaris}) 
find an age of $(9.2\pm1.0)$ Gyr for 47 Tuc.

The BSS population is clearly seen in the HR diagram 
in Figure \ref{fig:jcdmodel}.
The ``life span'' for models with the metallicity of 47 Tuc and
masses $1.4 \, M_\odot$ and $1.6 \, M_\odot$ (typical for BSS stars) 
is only about 1.8 Gyr and 1.3 Gyr, respectively (the time to reach the hook).
Thus it is evident that these stars have formed late in the
evolution of the cluster,
probably by a merger or collision between two stars
of mass below the turn-off mass.

The observed instability strip for {\em population I} $\delta$ 
Scuti stars is also
shown in Figure \ref{fig:jcdmodel} (Breger \cite{instab-strip}).
Four of five SX Phe stars are located inside 
the instability region. Even when taking into account the 
uncertainty of the temperature V14 is not inside the 
instability region --- possibly explaining the lack of variation 
for this BSS (cf.~the amplitude diagram in Figure \ref{fig:powerALL}).

Detailed modelling of the BSS stars is hampered by the uncertainties
relating to their formation (e.g. Benz \& Hills \cite{bss-merginge},
\cite{bss-merging}; 
Bailyn \& Pinsonneault \cite{bss-collision}).
This likely leads to changes in composition due to mixing;
in the extreme case where stars at the turn-off are completely
mixed, the resulting homogeneous helium content would be $Y \approx 0.34$.
Furthermore, it is plausible that, for example, the orbital angular
momentum in a merging binary star would lead to a very rapidly rotating BSS.
A detailed investigation of such effects is beyond the scope of the
present paper. 
However, we have explored the effects of a change in the composition:
In addition to the models with the 
standard helium abundance ({\em SHe}, $Y=0.241$),
we have considered models with increased helium content ({\em IHe}, $Y=0.261$)  
and masses $1.40 \, M_\odot$ and $1.50 \, M_\odot$;
these are also shown in Figure \ref{fig:jcdmodel} (dashed lines).

{}From the evolutionary tracks it is possible to estimate the masses
of the SX Phe stars. A unique case is V2 for which an unambiguous
estimate can be given:
The evolutionary tracks are almost horizontal around V2, 
hence the uncertainty in $T_{\rm eff}$ 
is not so important for the mass estimation from the position
in the HR diagram. For V2 we find $M_{\rm HR}/M_\odot\simeq1.54\pm0.05$ for
the standard helium abundance;
using instead the models with increased
helium abundance ({\em IHe}) we obtain $M_{\rm HR}/M_\odot=1.50\pm0.05$.
The other SX Phe stars all lie
around the hooks of the evolutionary tracks and the mass estimate
is more uncertain for these stars ($\pm0.10 \, M_\odot$). 
In Table \ref{tab:params} we have given mass estimates and
various parameters of the SX Phe stars.

In Figure \ref{fig:jcdmodel} the model ages in Gyr are given 
at three points along the $1.5 \, M_\odot$ track which is below V2
(in square brackets). Consequently the age of V2 must be 
somewhat lower than the $\approx 1.9$ Gyr found from this track.
If we assume that V2 is following the evolutionary track of a
standard model (ignoring formation history, e.g.~mixing of the
merging stars) and that the position in the HR diagram is correct
we can conclude (from an evolutionary track of mass $1.54M_\odot$): 
The age of V2 is around $(1.7\pm0.2)$ Gyr, $L/L_\odot = 17.0\pm0.5$, 
core burning of
hydrogen has ceased $(X_{\rm core} = 0)$, and the star has only 
a few hundred million years left before it will start climbing the 
Hayashi track and eventually go into the helium-burning phase.
\begin{table}[hbtp]

\caption{\label{tab:params}
Fundamental parameters 
for the observed SX Phe stars. 
Typical errors are $(0.03,0.04,0.14,0.02,200$~K$)$ for 
$(V,V-I,M_{\rm bol},{\rm BC}_V,T_{\rm eff}) $. For the mass in column 6 the error
estimate is $\pm0.05 \, M_{\odot}$ for V2 and $\pm0.10 \, M_{\odot}$
for the other variables}
\centering{
\begin{tabular}{ccccrrl} 
\hline
Star & $V$ & $V-I$ & $M_{\rm bol}$ & ${\rm BC}_V$ & $T_{\rm eff}$ & 
$M/M_\odot$  \\ \hline
 V2 &15.07  & 0.43 &  1.68 & -0.02 & 7100 & 1.54\\ 
 V3 &15.93  & 0.38 &  2.54 & -0.02 & 7300 & 1.35\\ 
V15 &15.28  & 0.24 &  1.91 & -0.01 & 7900 & 1.65\\ 
V16 &15.76  & 0.38 &  2.37 & -0.02 & 7300 & 1.35\\ \hline

\end{tabular}}
\end{table}

\subsection{Double-mode SX Phe stars}

For some double-mode SX Phe stars
a safe mode identification is possible.
Most double-mode pulsators oscillate in
radial modes ($l=0$) of low radial order ($n=1-6$).
One can put constraints on the astrophysical properties 
(mass, helium content) of these stars by comparing
observed and theoretical data in a period-ratio versus
period diagram, the so-called {\em Petersen diagram}.
In the following we
look more closely at the double-mode SX Phe star V2,  
and then discuss the more difficult stars V3, V15, and V16.

\subsection{\label{sec:dmvtwo}The double-mode SX Phe star V2}

We interpret the dominant observed modes as the 
the fundamental mode and the first overtone 
and obtain a period ratio of $\Pi_1/\Pi_0 = 0.7739\pm0.0002$.
Gilliland et al.~(\cite{gill98}) found $\Pi_1/\Pi_0 = 0.7722\pm0.0014$, 
i.e.~the results are essentially consistent. 
With the more accurate
period ratio obtained from the new longer time series we can
constrain the mass of V2 to within a formal error of 
$0.03 \, M_\odot$ at a given composition from the position 
in the Petersen diagram. The error in the mass estimate will be 
discussed further below.

\begin{figure}
 \hspace{-.7cm}
\resizebox{9.8cm}{!}
{\includegraphics{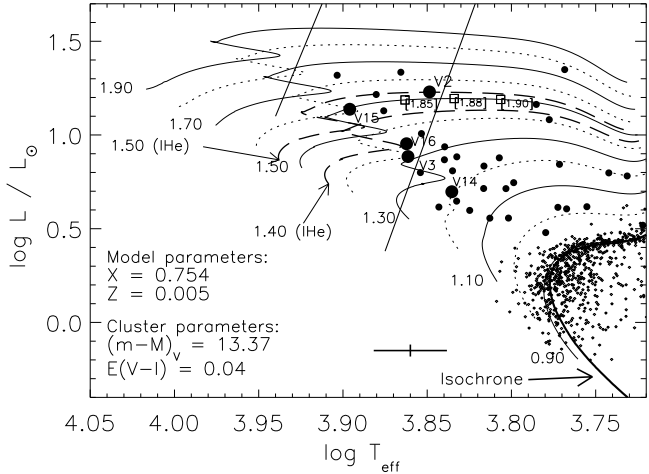}} 

\caption{HR diagram of 47 Tuc. Several theoretical evolutionary tracks
are shown. For models in the mass interval $M/M_\odot=0.9$ to $1.9$ 
and standard chemical composition tracks are plotted with either solid 
or dotted lines. 
Two tracks with increased helium content are also shown with dashed lines. 
For the $1.5 M_\odot$ model the age in Gyr is indicated at three points.
The filled circles mark the BSS stars and the SX Phe stars have been labeled.
The turn-off region can be seen in the lower right corner. 
The age of the isochrone is
12.5 Gyr (computed by VandenBerg, private communication)}
\label{fig:jcdmodel}
\end{figure}

We have used a series of
pulsation models which are extracted from the grid of evolutionary tracks
in Figure \ref{fig:jcdmodel}.
From the HR diagram we make a rough
estimate of the range of stellar masses that are relevant 
($1.8>M/M_\odot>1.3$). 
As before, in addition to the standard models we have also calculated 
pulsation models with higher helium content.

\begin{figure}
\resizebox{8.8cm}{!} 
{\includegraphics{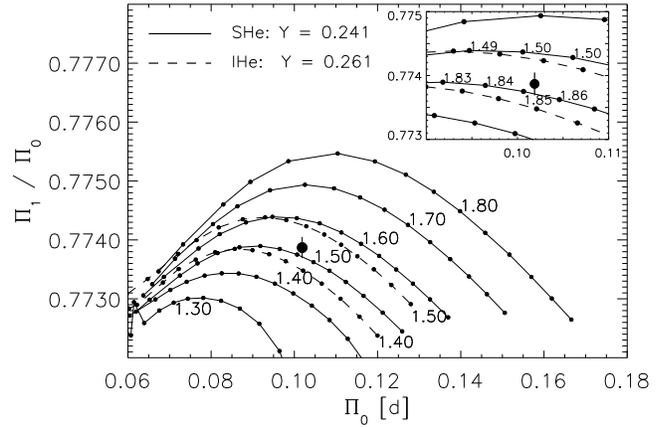}}
\caption{The Petersen diagram for several stellar model series. 
The helium content of the different model series is
explained in the top left corner;
as indicated, the {\em SHe} models cover the mass range
$M/M_\odot = 1.3 - 1.8$, while {\em IHe} models are shown for
$M/M_\odot = 1.4$ and 1.5.
In the inserted plot the ages of models with $Y=0.241$
and mass $M/M_\odot=1.5$ and 1.6 are given in Gyr}
\label{fig:petersen}
\end{figure}

Figure \ref{fig:petersen} compares the observed period ratio of V2 with data
for several stellar oscillation models. Solid lines are 
computed from models
with standard helium abundance ({\em SHe}, $Y=0.241$),
dashed lines have increased helium 
content ({\em IHe}, $Y=0.261$).
A decrease in effective temperature 
causes an increase in the pulsation period and as a consequence 
a stellar model moves from left to right in Figure \ref{fig:petersen} 
during its post-main-sequence evolution.

The double-mode SX Phe star V2 is shown in the plot. We estimate
the mass of V2 to be $(1.53\pm0.03) \, M_\odot$ for {\em SHe}.
From the models with {\em IHe} we infer 
the mass to be $(1.45\pm0.03) \, M_\odot$.

The good accuracy of the observed period ratio for V2
--- the error bars can be seen in the inserted plot ---
is due to the long time series with good sampling.
In comparison the result by Gilliland et al.~(\cite{gill98}) is 
$\left(1.4^{+0.25}_{-0.15}\right) \, M_\odot$ (cf.~their Figure 22).


The mass determination is obviously also sensitive to errors in
the stellar modelling used in computing periods and period ratios.
We have estimated the intrinsic errors arising from numerical
effects in the computation of the evolutionary models and their
oscillation frequencies; the effect on the computed period ratio $\Pi_1/\Pi_0$
is minute, less than $10^{-4}$.
On the other hand,
major contributions to the uncertainty in the mass determination arise from
uncertainties in the composition parameters and possible effects from
rotation of V2. 
The possible effects on the helium abundance of mixing associated
with the formation of BSS have already been mentioned.
To this must be added the uncertainties in the determination of
the overall composition of the cluster.
Also, it is well known that rotation affects the evolution of stellar models
and the derived oscillation frequencies (e.g.~Pamyatnykh 2000,
Goupil et al.~2000). Fast rotation modifies the evolutionary tracks in
the HR and Petersen diagrams, particularly in the post-main-sequence
stages where V2 is located. Although the frequency corrections are
expected to be relatively small for radial modes, they may be significant.
For the high-amplitude $\delta$ Scuti stars 
(with amplitude $A_V > 0.3$ mag) rotation is usually slow.
However, V2 has relatively low amplitude ($A_V = 0.15$) 
and its rotational velocity ($v\sin i$) is not known.
Finally, we recall from Section \ref{sec:cmd} that basic
parameters of the star, i.e. the effective temperature and
the distance modulus, are subject to substantial uncertainty.

Given these uncertainties, to what extent can we trust our mass determinations?
A very significant aspect in the interpretation is the {\it consistency}
between different determinations.
As an example, consider the effects of the helium abundance,
exemplified by the {\em SHe} and {\em IHe} model series.
{}From the analysis of the HR diagram (Figure \ref{fig:jcdmodel}) 
V2 has {\em evolutionary mass} $M_{\rm HR}\simeq (1.54\pm 0.05) \, M_\odot$ with
{\em SHe}, 
for which the period-ratio mass is $(1.53 \pm 0.03) M_\odot$;
for {\em IHe} the corresponding values are
$M_{\rm HR}\simeq (1.50\pm0.05) \, M_\odot$ and 
$(1.45 \pm 0.03) M_\odot$ from the period ratio.
Thus for both compositions the two mass determinations
agree within the formal error bars, although the match
is somewhat better for {\em SHe}.
We also note that the mass estimates allow a better
constraint on the change in the helium abundance resulting from complete mixing.
Assuming that the progenitor stars had masses of around 
$0.75 M_\odot$ and had evolved for 10 Gyr before the merger,
the average helium abundance would be increased by around 0.04
relative to the original value, i.e.~twice the difference
between {\em SHe} and {\em IHe}.
Since complete mixing is thought to be unlikely during BSS formation,
the {\em SHe} abundance solution is favoured ($Y = 0.241$).

We have adopted a metal content $Z =  0.005$ for the
models. However, the value of ${\rm [Fe/H]} = -0.83$ from Table 1 with the
original definition ${\rm [Fe/H]} = \log (Z/Z_{\odot})$
and $Z_{\odot} = 0.019$
gives $Z =  0.0028$, and Salaris \& Weiss (1998) use $Z=0.008$,
since they account for $\alpha$-element enhancement. 
In addition to the analysis based on 
standard stellar model series with $Z =  0.005$, we have also
performed mass estimations using the composition used by 
Salaris \& Weiss (\cite{salaris}), i.e.~$Y=0.273, Z= 0.008$. 
We find that for this composition we cannot obtain consistent 
modelling satisfying the observational constraints: 
V2 fits an evolutionary track in the HR diagram of mass $1.56 \, M_\odot$ 
and the period ratio yields the mass $1.67 \, M_\odot$.

\begin{table}
\caption[ ]{
Mass determination of the SX Phe star V2 from its position in
the HR diagram and from comparison of observed
modes with pulsation models. The results when 
using different chemical compositions are given (see text for details)}

\label{massofv2}

\vskip 0.2cm

\centering{
\begin{tabular}{c|ccc}  \hline

	      Composition & {\em SHe} & {\em IHe} & {\em S-W}\\ \hline
              $Y$           & 0.241 & 0.261 & 0.273 \\
	      $Z$           & 0.005 & 0.005 & 0.008 \\ \hline
Method  &  \multicolumn{3}{c}{Mass estimates ($M/M_\odot$)} \\ \hline
 	      HR-diagram  & 1.54  & 1.50  & 1.56 \\
        Pulsation modes   & 1.53  & 1.45  & 1.67 \\ \hline

\noalign{\smallskip}

\end{tabular}}
\end{table}

From the period and period ratio we may also infer the
age and effective temperature of V2.
For the {\em SHe} models of mass $1.6$ and $1.5 \, M_\odot$ we have
indicated the age of the models in Gyr in the inserted plot in
Figure~\ref{fig:petersen};
this can be compared with the ages in the HR diagram 
in Figure~\ref{fig:jcdmodel} for the $1.5 M_\odot$ model. 
The analysis indicates that $T_{\rm eff}$
may be somewhat higher than inferred from our adopted $V-I$ photometry:
From the Petersen diagram we infer that $T_{\rm eff} = (7400\pm150)$~K.
The ages roughly agree with Figure \ref{fig:jcdmodel} when taking the
error on $T_{\rm eff}$ into account, leading to an age of $(1.7\pm0.1)$ Gyr.

The determination of $M_{\rm HR}$ is sensitive to errors
in the distance modulus..
An independent estimate of this may be obtained from the
period-luminosity-colour-metallicity (PLCZ) relation for
SX Phe stars, derived by
Petersen \& Christensen-Dalsgaard (\cite{otzen99}) from
theoretical stellar models.
This relation is sensitive to the adopted values of $Z$ and in
particular of $T_{\rm eff}$. 
In the above analysis of V2 we have considered the $T_{\rm eff}$-interval
$7100$ to $7700$~K. For the preferred $Z = 0.005$ the PLCZ relation gives
the corresponding interval in $M_{\rm bol}$ of 1.98 to 1.56 mag. We note that
this is in rough agreement with the value given 
in Table 3 for V2 ($M_{\rm bol}$ = 1.68)
and hence justifies our use of the VandenBerg (\cite{vandenberg2000})
distance modulus.
Although this is somewhat speculative, a
higher temperature for V2, as inferred from the
period, seems to be necessary to explain the distance 
modulus inferred from the main period. 
We recall from Section \ref{sec:cmd} that independent estimates of 
$V-I$ gave discrepant results for V2.

We thus have two clues that indicate the $T_{\rm eff}$ of V2 may be
higher than what we find from the colour
(i.e.~$V-I = 0.43$ or $T_{\rm eff}=7100$~K): 1) A comparison of the
theoretical model ages that best fit the position in the
Petersen diagram and the HR diagram do not agree. 2) The PLCZ calibration from
Petersen \& Christensen-Dalsgaard (\cite{otzen99}) gives a small
distance modulus ($(m-M)_V=13.1\pm0.1$)
for $Z = 0.005$ and $T_{\rm eff} = 7100$~K.
Both these observations indicate a higher value of $T_{\rm eff}$.
If for V2 we assume $T_{\rm eff}=(7500\pm200)$~K
appropriate to the independent $V-I$ = 0.27 determination 
we find that the PLCZ relation gives a distance modulus
of $(m-M)_V=13.4\pm0.1$
in agreement with VandenBerg (\cite{vandenberg2000}) 
and Grundahl et al.~(\cite{grundahl}) --- and with this temperature
the new position of V2 in the HR diagram agrees with
the position in the Petersen diagram within error-bars.

It is evident that there are a number of uncertainties concerning
the parameters and modelling of V2.
Also, we have not been able to take into account the possible
effects of rotation on the inferences.
However, the fact that
we obtain a consistent description of V2 using standard modelling
without rotation --- which is the simplest possibility ---
gives us some confidence in this modelling. 
Since V2 is a relatively evolved and hence old BSS at $\sim$1.7 Gyr
it could by now have lost significant angular momentum even if it
was formed initially as a rapid rotator.
Thus, from this discussion 
we maintain that a reasonable estimate for the mass of V2 is
$M/M_\odot = 1.54\pm0.05$; we stress, however, that this assumes V2
to have the same chemical composition as an average star in the cluster.

In Table \ref{massofv2} we present the results of the mass estimation 
of V2 for the three different chemical compositions discussed above.

We finally note that we have also found new low-amplitude modes for V2
(see Table 2), but their frequencies
are simply linear combinations of the frequencies of the two main modes. 
This indicates that the modes are interacting
non-linearly as was also noted by Gilliland et al.~(\cite{gill98}), and the
pattern given in Table 2 is in perfect agreement with predictions from Garrido
and Rodriguez (1996).

\subsection{V3, V15, and V16}

The interpretation
of the modes of V3 is difficult since the period ratio 
is close to unity. Gilliland et al.~(\cite{gill98}) propose that it is
a rotationally split non-radial mode ($l=1$). If this is true the
rotational period of V3 (based on the splitting in Table \ref{tab:periods})
is $P_{\rm rot}=(2.33\pm0.03)$ days. Gilliland et al.~(\cite{gill98}) found
a significantly faster rotation period $P_{\rm rot} = (1.65\pm0.2)$ days.

\begin{figure}
\resizebox{8.8cm}{!}
{\includegraphics{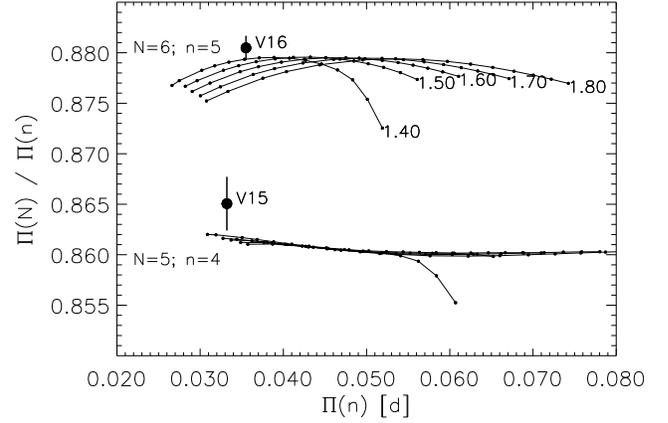}}
\caption{Petersen diagram showing radial overtones of orders four to six
for several stellar model series. 
The $\bullet$ mark the SX Phe stars V15 and V16. The vertical bars are
the 10$\sigma$ error on the period ratio}
\label{fig:petersen-v15}
\end{figure}

The interpretation of the double-mode SX Phe 
star V15 is also difficult. If this star is assumed to be a ``classical''
double-mode pulsator 
oscillating in radial modes, it must be oscillating in high overtones.
The observed period ratio is $\Pi_N/\Pi_n = 0.8651\pm0.0003$ which
is below the ratio calculated from theoretical models 
with radial orders 5 and 6 --- but above the calculated ratio for
radial orders 4 and 5.
Figure \ref{fig:petersen-v15} 
shows several series of models oscillating in these high overtones.
The 10$\sigma$ error bars are shown.

The result for V15 by Gilliland et al.~(\cite{gill98})
was $\Pi_N/\Pi_n = 0.8698\pm0.0023$. This result
was consistent with V15 oscillating in two high overtones 
with their analysis, i.e.~orders 4 and 5.
With the new results none of the
calculated models fits the observed properties of V15.
This is also the case when using the composition from Salaris 
\& Weiss (\cite{salaris}), i.e.~models with higher helium 
and heavy element content.

Possible explanations for the apparent discrepancy are e.g.~(i) that 
at least one of the observed modes is a non-radial mode, 
or (ii) that the frequencies are modified by rotation of the star.
A clear interpretation of the modes in V15 cannot yet be made.

We note that for the two modes in V15 the periods 
determined from 1993 and 1999 observations
agree to 0.8$\sigma$ and 2.0$\sigma$ for the large- and small-amplitude
peaks, respectively.  
Either spurious agreement with radial overtone period ratios was allowed by 
the larger errors in the results of Gilliland et al.~(\cite{gill98}),
or the lower-amplitude peak
represents different modes having been excited 6 years apart.

Interpretation of the double-mode star V16 is also problematic here.
The mode near 32 cycles/day from 1993 is clearly present in the 1999
observations with the new frequency 1.2$\sigma$ from the original value.
The mode near 28 cycles/day is at a lower amplitude in 1999 and at
marginal $\sim$3$\sigma$ significance in each epoch separately.
Taking the agreement (to 0.8$\sigma$ from the 1999 period) as
confirmation of this mode and evaluating the period ratio yields
$\Pi_N/\Pi_n = 0.8805\pm0.0002$ (compared to 1993 data result
of 0.8816 $\pm$ 0.0028) which falls just above the period
ratios for orders 5 and 6 in Figure 12.
However, there are evidently other modes present to equal or higher
significance -- these cannot all be radial modes.  The significance of 
amplitudes quoted in Table 2 represent very conservative values
based upon establishing a {\em local} mean in the amplitude spectra
and considering the ratio of the mode amplitude to this.
Since our power spectra for non-variable stars remain flat consistent
with white noise, it may well be appropriate to base the errors
on means over the full available frequency range.
As an example, for V16 adopting a global normalization for the 
amplitude spectrum noise would increase the significance of the 
three modes near 28, 29, and 31 cycles/day to over 8$\sigma$ each
with the possibility that yet weaker modes are also present.

\subsection{Future work}

How can the analysis
of the SX Phe variables in 47 Tuc be improved in
the future? From the discussion above it is evident that accurate
determinations of effective temperatures will give stronger constraints
on the modelling. Secure determinations of the chemical composition and
studies of rotation will also be important.

A more ambitious project is simultaneous 
modelling of several SX Phe variables in 47 Tuc following the philosophy
of the programme for asteroseismology in open clusters 
(e.g.~Frandsen et al.~1996). 
The new results for V15 and V16 are consistent with the presence of
some non-radial modes, perhaps with high overtone radial modes
present as well.
Another challenging case is V1, 
which Gilliland et al.~(\cite{gill98}) found to be a 
triple-mode variable with period ratios
that cannot be explained by
radial oscillations of standard evolutionary models with the
accepted composition of 47 Tuc.
Either this star, with an amplitude of about 0.15 mag, has a low $Z \approx 0.001$, 
its modes are non-radial or perhaps a large effect from rotation needs 
to be considered. It seems clear that more detailed studies of the SX Phe
stars in 47 Tuc will provide valuable asteroseismological information.

\section{Conclusions}

The unique time series of 8.3 days from HST of 47 Tuc 
analyzed here was optimized to look for planets around 
stars fainter than the turn-off. 
We have shown that it is possible to obtain good photometry
for the saturated stars by identifying and flagging (i.e.~not using) pixels 
that are contaminated by neighbouring pixels due to {\em bleeding} signal. 
We have thus extracted good (1-3\% noise) light curves
for the saturated stars. No new variable BSS stars were 
found in the sample but have analyzed five of the six known
SX Phe stars in the core of 47 Tuc (V1 was not in the field of view).
For V14 we do not detect the
oscillation signal that has been claimed previously. 
For three of the double-mode stars (V2, V15, and V16) 
we have used theoretical stellar
models to attempt to determine their masses: Both through the position in the
HR diagram (using the observed magnitudes) and in the Petersen diagram 
(using the observed periods).

The most striking result from this study is the determination of
the mass of the SX Phe star V2. From the HR diagram 
(not depending on the uncertainty of $T_{\rm eff}$) and the
Petersen diagram we infer a mass of $M_{\rm V2} = (1.54\pm0.05) \, M_\odot$.
Important sources of error in this mass
estimate of V2 are the chemical composition 
and effects of rotation;
even so, we believe that we have obtained
the so far most precise determination of the mass of a 
BSS (see Shara et al.~\cite{shara}).
Further progress requires
a spectroscopic study to constrain $T_{\rm eff}$, rotational velocity, 
and chemical composition for V2;
this would be difficult from the ground due to crowding, 
but it is feasible with HST. 
Our results indicate that,
with such additional data, we would be able to obtain strong 
constraints on the processes that lead to the formation of BSS.

\begin{acknowledgements}
 We wish to express our gratitude to Don VandenBerg for providing us
 with a set of independently calculated evolutionary models and an 
 isochrone for 47 Tucanae.
 This research was supported by the Danish Natural Science Research 
 Council through its Centre for Ground-based Observational
 Astronomy and by the Danish National Research Foundation
 through its establishment of the Theoretical Astrophysics Center. 
U.S. investigators were supported in part by STScI Grant
GO-8267.01-97A to the Space Telescope Science Institute and several
STScI grants from the same proposal to co-I institutions.
\end{acknowledgements}

\end{document}